\documentclass[a4paper,11pt]{article}

\usepackage{jheppub} 

\usepackage[T1]{fontenc} 

\title{The collision of two-kinks defects}

\author[a]{T. S. Mendon\c ca}

\author[a,b,1]{H. P. de Oliveira \note{Corresponding author.}}

\affiliation[a]{
Departamento de F\'{\i}sica Te\'orica - Instituto de F\'{\i}sica
A. D. Tavares\\ Universidade do Estado do Rio de Janeiro, 
Rio de Janeiro, RJ, 20550-013, Brazil}

\affiliation[b]{Department of Physics and Astronomy \\
        University of Pittsburgh, Pittsburgh, PA 15260, USA}

\emailAdd{tiagobrouwer@msn.com}
\emailAdd{hp.deoliveira@pq.cnpq.br}

\abstract{
We have investigated the head-on collision of a two-kink and a two-antikink pair that arises as a generalization of the $\phi^4$ model. We have evolved numerically the Klein-Gordon equation with a new spectral algorithm whose accuracy and convergence were attested by the numerical tests. As a general result, the two-kink pair is annihilated radiating away most of the scalar field. It is possible the production of oscillons-like configurations after the collision that bounce and coalesce to form a small amplitude oscillon at the origin. The new feature is the formation of a sequence of quasi-stationary structures that we have identified as lump-like solutions of non-topological nature. The amount of time these structures survives depends on the fine-tuning of the impact velocity.}

\begin{document} 
\maketitle
\flushbottom

\section{Introduction}

Topological defects are of great interest due their ubiquity in many branches of physics like fluid mechanics, condensed matter, nuclear and particle physics and cosmology. The simplest form of topological defects are the one-dimensional kinks modeled by a scalar field $\phi$ satisfying the Klein-Gordon equation,
\begin{equation}
\frac{\partial^2 \phi}{\partial t^2} - \frac{\partial^2 \phi}{\partial x^2} + \frac{\partial V(\phi)}{\partial \phi} = 0, \label{eq1}
\end{equation}
\noindent where $V(\phi)$ is the scalar field potential. A large variety of kink solutions arises for potentials with at least two distinct minima. The most common models of kinks \cite{rajaraman,vachaspati} are described  by the potential $V(\phi)=(1-\phi^2)^2/2$ that defines the $\phi^4$ model, and $V(\phi) = 1-\cos(\phi)$ known as the sine-Gordon model. 

The collision of kinks has been studied in details motivated mainly by the applications many areas of physics. The nonlinear nature of the problem governed by a time-dependent partial differential equation is responsible for new and unexpected features. We mention the fractal structure in the collision of a kink and an anti-kink in the $\phi^4$ model \cite{sugiyama,moshir,fractal1,fractal2,aninos,goodman}, and also found in other models such as the parametrically modified sine-Gordon \cite{sine_mod,sine_mod2} and the $\phi^6$ models \cite{phi6_2010,phi6,phi6_2014}. Remarkably, the exception is provided by the collision of sine-Gordon kinks in which the Klein-Gordon equation is solved exactly.

Some years ago Bazeia et al \cite{bazeia} proposed a new class of topological defects in systems described by real scalar fields in $(D,1)$ spacetime dimensions. By restricting to $D=1$ they have considered the family model with potential,
\begin{equation}
V(\phi) = \frac{1}{2} \phi^2 (\phi^{-1/p}-\phi^{1/p})^2, \label{eq2}
\end{equation}
\noindent where the parameter $p$ is related to the way the scalar field self-interacts. This model can be understood as a generalization of the $\phi^4$ model, which is recovered for $p=1$. For $p$ even, the case $p=2$ is special and describes an unstable lumplike configuration \cite{bazeia}. For $p=4,6,..$ the potential (\ref{eq2}) requires that $\phi \geq 0$ or $\phi \leq 0$ under the change $\phi \rightarrow -\phi$. As pointed out by Refs. \cite{bazeia,bazeia_deformed} it is possible to construct topological defect in the form $\phi^{(+)}=\tanh^p(x/p)$ ($x \geq 0$) and $\phi^{(-)}=-\tanh^p(x/p)$ ($x \leq 0$). Moreover, non-topological lumplike defects can be envisaged according to Ref. \cite{new_lumplike}. It can be shown that all defects with $p$ even, topological or non-topological, are unstable and therefore of few interest.

The new static structures arising when $p=3,5..$ connect the minima $\phi=\pm 1$ passing through the symmetric minimum at $\phi=0$ are called two-kinks defects \cite{bazeia}. As shown in Fig. 1, the two-kinks defects seem to be composed of two standard kinks symmetrically separated by a distance proportional to $p$. The energy density profile reinforces this view. However, Uchiyama \cite{uchiyama} presented for the first time a two-kinks structure after extending a hadron model. For the sake of completeness, the boosted two-kinks defects are described by the following exact solution,
\begin{equation}
\phi_{K,\bar{K}}(x,t) = \pm \tanh^p\,\left(\frac{x-x_0-u t}{p \sqrt{1-u^2}}\right), \label{eq3}
\end{equation}
\noindent where $\pm$ denotes the two-kink ($K$) and the two-antikink ($\bar{K}$), respectively. The parameter $u$ stands for the velocity of the kink along the axis and $x_0$ the center of the two-kink/anti-kink. 

The total conserved energy $E$ associated to the scalar field is calculated from,

\begin{equation}
E = \int_{-\infty}^{\infty}\,\rho_\phi(x,t) dx, \label{eq4}
\end{equation}

\noindent where the energy density is given by,

\begin{equation}
\rho_\phi(x,t) = \frac{1}{2} \left(\frac{\partial \phi}{\partial t}\right)^2 + \frac{1}{2} \left(\frac{\partial \phi}{\partial x}\right)^2 + V(\phi). \label{eq5}
\end{equation}

\noindent It can be shown that $E = 4p/(4p^2-1)\sqrt{1-u^2}$ is the conserved total energy of the boosted two-kink/anti-kink. 

It is worth pointing out that the two-kink defects can describe the magnetic domain walls in constrained geometries \cite{dw_exps,dw_exps2}. We also mention possible applications of two-kinks structures in the brane-world scenario \cite{chumbes,dutra}, and in connection with processes of their formation in perturbed sine-Gordon models \cite{monica}.

\begin{figure}[ht]
\begin{center}
\includegraphics[width=5.5cm,height=4cm]{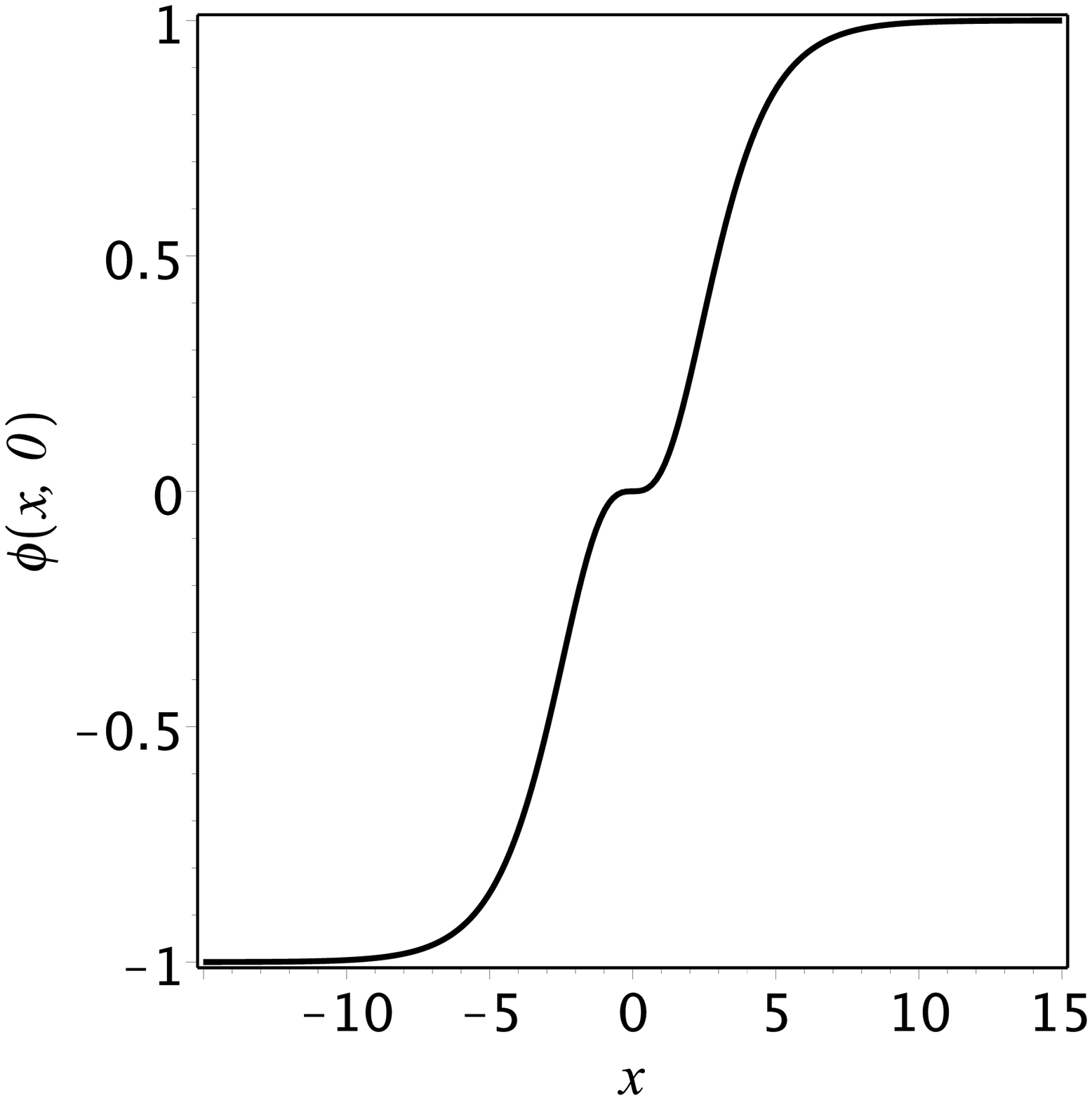}
\includegraphics[width=5.5cm,height=4cm]{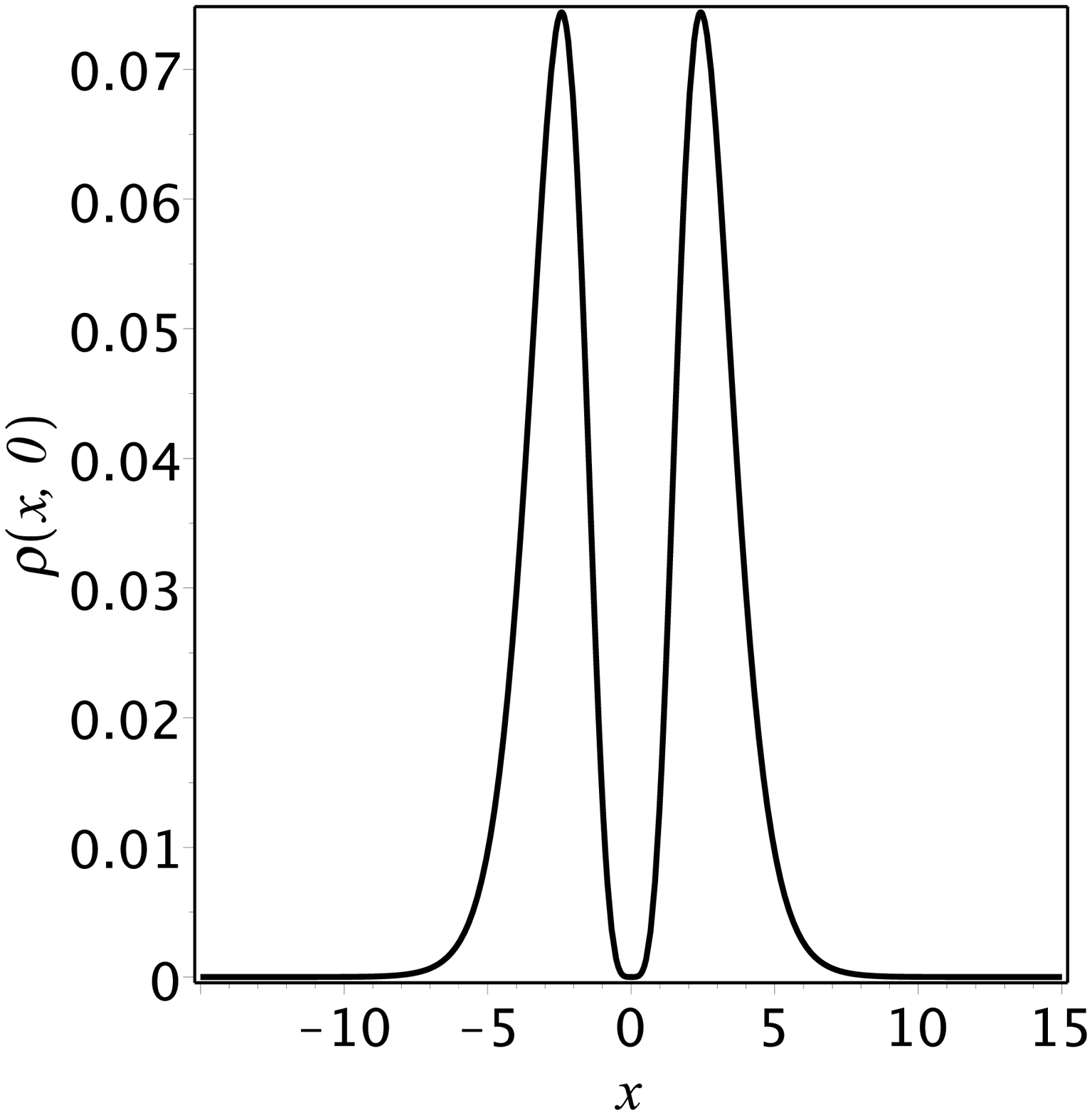} 
\end{center}
\caption{Profiles for the kink and its energy density for $p=3$ and $u=0.4$. The resulting structures are composed of two standard kinks symmetrically separated by a distance proportional to $p$ \cite{bazeia}.}
\end{figure}

The main purpose of this paper is to study numerically the collision of a two-kink-two-antikink pair with a code based on spectral methods. The paper is organized as follows. In Section 2 we have introduced the numerical method that can be applied to any one-dimensional scalar field model. We have performed numerical tests showing the accuracy and convergence of the algorithm considering the exact breather solution of the sine-Gordon model, and the collision of standard kinks of the $\phi^4$ model. In Section 3 we have proceeded by studying the head-on collision of a two-kinks  pair for $p$ odd. The outcomes are quite distinct from the collision of kink-antikink pair of the $\phi^4$ model. In Section 5, we have summarized the results and discussed future perspectives of the present work. 

\section{The numerical method}

The Klein-Gordon equation (\ref{eq1}) can be solved numerically using straightforwardly the Collocation method \cite{boyd} as we describe in the sequence. The first step is to establish the approximate scalar field as,
\begin{equation}
\phi_a (t ,x)=\sum_{k=0}^N\,a_k(t) \psi_k(x),  \label{eq6}
\end{equation}
\noindent where $a_k(t)$, $k=0,1,..,N$ are the unknown coefficients, the functions $\psi_k(x)$ constitute the basis functions chosen suitably as a generalization of a Fourier expansion, and $N$ is the order of the series truncation. It remains, therefore, to compute the $N+1$ unknown coefficients $a_k(t)$ to complete the approximate solution.

The solutions describing the dynamics of kinks are defined in the unbounded domain $(-\infty,+\infty)$ and obtained after solving a nonlinear partial differential equation. In this vein, we mention two possible strategies for treating numerically this problem. The first is to consider the finite difference scheme using staggered leapfrog integration \cite{press} of the equation of motion in a finite spatial domain with periodic boundary conditions. Pseudospectral method was used in a finite domain with periodic boundary conditions for the $\phi^6$ model \cite{phi6}. The second strategy which we will adopt here is to map the infinite interval to a finite domain. According to Boyd \cite{boyd}, it is possible to generate new basis functions for the infinite interval as images under the change of coordinate of, for instance, Chebyshev polynomials or Fourier series. There is a large variety of maps, some of them are more popular than others. In order to decide which map is more appropriate for the problem under consideration, we recall that a typical kink exact solution decays as powers of the hyperbolic tangent as $|x| \rightarrow \infty$. In this case, we have found that the logarithmic map \cite{boyd,boyd95a,gotlieb} is well suited for the problem under consideration. The logarithmic map is given by, 
\begin{eqnarray}
\eta=\tanh\left(\frac{x-\bar{x}_0}{L_0}\right), \label{eq7}
\end{eqnarray}    
\noindent where $\eta \in [-1,1]$, $L_0$ is the map parameter and $\bar{x}_0$ defines the origin of the computational variable $\eta$. We can define the basis functions as the Chebyshev polynomials, $T_k(\eta)$, or their images under the map (\ref{eq7}) defined on the physical domain $x \in (-\infty,\infty)$:
\begin{equation}
\psi _k(x) \equiv T_k\left(\tanh\left(\frac{x-\bar{x}_0}{L_0}\right)\right). \label{eq8}
\end{equation}

In what follows the collocation method is applied straightforwardly. The first step is to substitute the approximate scalar field into the Klein-Gordon equation to form the residual equation,
\begin{eqnarray}
\mathrm{Res}_{KG}(t,x) &=& \sum_{k=0}^N\,\dot{v}_k(t) \psi_k(x) - \sum_{k=0}^N\,a_k(t) \psi_k^{\prime\prime}(x) \nonumber \\
&&+ \left(\frac{d V}{d \phi}\right)_{\phi=\phi_a}, \label{eq9}
\end{eqnarray}  
\noindent where we have introduced the `velocities' $v_k = \dot{a}_k$. Next, the unknown coefficients are determined with the condition of vanishing the residual equation at the collocation or grid points on the physical domain $x_j$, 
\begin{equation}
\mathrm{Res}_{KG}(t,x_j) = 0, \label{eq10}
\end{equation}
\noindent for all $j=0,1,..,N$. The collocation points $x_j$ are determined from the map (\ref{eq7}) taking the collocation points defined on the computational domain,
\begin{equation}
\eta_j = \cos\left(\frac{j\pi}{N}\right),\;\; j=0,1,..,N. \label{eq11}
\end{equation}
\noindent Therefore, we ended up with a set of $N+1$ first order ordinary differential equations for the velocities $v_k$, which must be complemented with the set of $N+1$ differential equations, $\dot{a}_j=v_j$, to complete the whole set of equations for the unknown coefficients. While the first two terms of the residual equation can be expressed spectrally, that is, in terms of the unknown coefficients, the last term is treated using the physical representation. It means that instead of the spectral representation of the scalar field through the coefficients $a_k(t)$, we have considered the physical representation using the values of the scalar field at the collocation points, $\phi_j(t)=\phi_a(t,x_j)$. Both representations are related since,
\begin{equation}
\phi_j(t)=\phi_a(x_j,t)=\sum_{k=0}^N\,a_k(t) \psi_k(x_j). \label{eq12}
\end{equation}
The resulting dynamical equations are,
\begin{eqnarray}
\dot{a}_k &=& v_k \label{eq13} \\
\dot{v}_k &=& F_k(a_j,\phi_j), \label{eq14}
\end{eqnarray}
\noindent where the first set relates the definition of the velocities and the second set arises after solving the relations represented by (\ref{eq10}). The above dynamical system is complemented with $N+1$ algebraic relations,
\begin{equation}
\phi_j = \phi_j(a_k). \label{eq15}
\end{equation}
\noindent To integrate the above equations one needs the initial data:
\begin{equation}
\phi(0,x)=f(x),\;\; \left(\frac{\partial \phi}{\partial t}\right)_{t=0}=g(x), \label{eq16}
\end{equation}
\noindent from which the initial coefficients $a_k(0)$ and $v_k(0)$ can be calculated.

\begin{figure}[ht]
\begin{center}
\includegraphics*[width=6.cm,height=4.5cm]{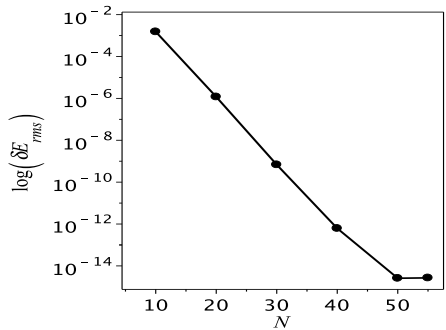} 
\end{center}
\vspace{-0.5cm}
\caption{Exponential decay of $\delta E_{\mathrm{rms}}$ evaluated from $t=0$ to $t_f=70$, after setting $\omega=0.5$ and $t_0=1.0$ into the exact solution (\ref{eq15}), and $L_0=10$ for the map parameter. The saturation due to round-off error is achieved for $N \geq 50$.}
\end{figure}

\begin{figure}[htb]
\begin{center}
\includegraphics*[width=5.5cm,height=5cm]{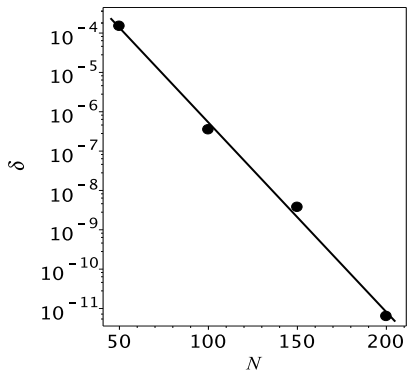} 
\end{center}
\vspace{-0.5cm}
\caption{Exponential decay of $\delta E_{\mathrm{rms}}$ for the collision of a kink and an antikink in $\phi^4$ model for $u=0.3$.} 
\end{figure}

The classical sine-Gordon model offers exact solutions describing the interaction of kinks initially at rest or moving towards each other \cite{vachaspati}. We have considered the exact solution known as the breather to serve as a benchmark to our numerical algorithm. This solution describes the interaction of a pair of kink/anti-kink interacting to form a periodic bound state, and is given by,
\begin{equation}
\phi(t,x) = 4 \arctan \left[\frac{\sqrt{1-\omega^2} \sin(\omega(t+t_0))}{\omega \cosh(\sqrt{1-\omega^2} x)}\right], \label{eq15}
\end{equation}   
\noindent where $\omega$ is the frequency of the breather. We have determined the initial data (\ref{eq14}) corresponding to the above solution and evolve the dynamical equations (\ref{eq11}) and (\ref{eq12}). Instead of comparing the exact solution with the approximate solution given by (\ref{eq15}), we have used the deviation from the conserved total energy $E$ as the error control. More specifically, after evaluating the total energy associated to the numerical solution at each instant, $E(t)$, we determined the quantity,
\begin{equation}
\delta E(t) = \frac{E_{\mathrm{exact}}-E(t)}{E_{\mathrm{exact}}}, \label{eq16}
\end{equation}
\noindent where $E_{\mathrm{exact}}$ is the conserved exact total energy. We have set $\omega=0.5$, $t_0=1.0$ and evolved numerically the KG equation from $t=0$ to $t=70$ for increasing truncation orders, namely $N=10,20,..,50$. In order to present a clear effect of increasing the truncation order, we have plotted the rms of $\delta E$; $\delta E_{\mathrm{rms}}$, in terms of $N$. The result is presented in Fig. 2 with the expected exponential decay of $\delta E_{\mathrm{rms}}$ until the saturation due to round-off error is achieved for $N \geq 50$. 

We have performed an additional numerical test of checking the conserved total energy of the collision of a pair of kink-antikink in the $\phi^4$ model ($p=1$). The initial configuration is given by $\phi(x,u_0) = -1 + \tanh\left(\frac{x+x_0-ut}{\sqrt{1-u^2}}\right) - \tanh\left(\frac{x-x_0-ut}{\sqrt{1-u^2}}\right)$, where we have set $x_0=12$ and fixed the boost parameter as $u=0.3$. We have evolved the field equations (\ref{eq11}) and (\ref{eq12}) with the truncation orders $N=50,100,150,200$ and evaluated, in each case the error rms in the energy conservation from $t=0$ to $t=100$ when both kinks have collided and moved apart each other. Fig. 3 depicts the results. Accordingly, even for a modest truncation order of $N=50$, the energy is preserved to about a one part in $10^3$, whereas for $N=200$ the conservation of energy achieves about less than one part in $10^7$. 

\section{Collision of two-kinks defects}

\begin{figure}[htb]
\begin{center}
\includegraphics*[width=5.cm,height=4.5cm]{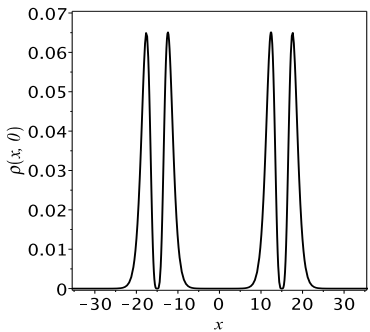} 
\end{center}
\caption{Illustration of the initial distribution of the energy density $\rho(x,0)$ corresponding to a pair of a two-kinks with impact velocity $u=0.2$ described by Eq. (\ref{eq17}). The two-kinks are placed at $x_0=15$, and this initial distribution is similar to four kinks of $\phi^4$ model.}
\end{figure}

\begin{figure}[htb]
\begin{center}
\includegraphics*[width=10.cm,height=8.5cm]{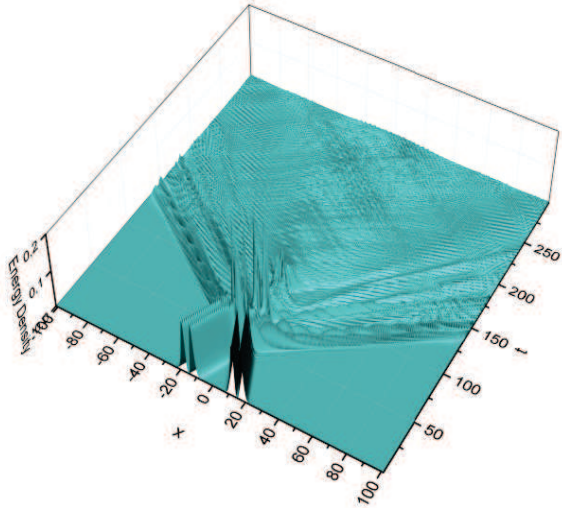}
\end{center}
\vspace{-0.5cm}
\caption{Three-dimensional plot of the energy density for the collision of two-kinks defects for $u=0.3$. Both two-kinks interacts in a very intricate way resulting in their complete disruption. The scalar field is radiated away leaving almost no structure about the origin}.
\end{figure}

As we have mentioned in the Introduction, the collision of a kink and anti-kink in the $\phi^4$ model exhibits a complex structure that depends on the impact velocity. This complex structure is expressed by the fractality associated to the transition between two possible outcomes: the formation of a bounded oscillating configuration and reflection of the kinks after the collision. These outcomes arise when the impact velocity is smaller or greater than a critical velocity, $v_c$, respectively. However, for some values of the impact velocity below $v_c$ the kinks become trapped, but it appears an intricate pattern of resonant windows in which they can escape to infinity. The accepted explanation for this feature is a nonlinear competition or transfer of energy between the translational and vibrational modes of individual kinks \cite{fractal2}. 

We describe here the head-on collision of a two-kink and two-antikink pair for $p \geq 3$. The starting point is to establish the initial data representing both two-kinks configurations moving initially toward each other with velocity $u$. The initial data is given by,
\begin{equation}
\phi_0(x,0) = -1+\phi_K(x+x_0,0) - \phi_K(x-x_0,0), \label{eq17}
\end{equation}
\noindent where we have chosen $x_0$ as the initial distance of the two-kinks defects. It is interesting to point out that the collision of both two-kinks might seem as the collision of four standard kinks. We show in Fig. 4 the initial profile of the energy density $\rho(x,0)$. 

We begin with $p=3$ that represents the most immediate generalization of the $\phi^4$ model ($p=1$), however most of the main features are also found on $p \geq 3$. We have performed numerical experiments with the velocity $u$ as the free parameter. In all simulations, we have considered $N=250$ and taking into account only the even modes in the spectral expansion (\ref{eq3}) due to the symmetry of the problem. We have also monitored the error measured by the relative deviation of the total energy given by Eq. (\ref{eq16}). It is worth mentioning that the error is sensitive to the choice of the map parameter $L_0$ (cf. (\ref{eq7})). Thus, after some trial and error we have fixed $L_0=90$ such that the error does not surpass $0.01\%$. We have considered acceptable, in spite of being much greater than the typical $10^{-9}\%$ obtained for the collision in the $\phi^4$ model. We attribute this discrepancy to the nature of the outcomes arising after which we are going to describe.  

\begin{figure}[htb]
\begin{center}
\includegraphics*[width=11.cm,height=9.cm]{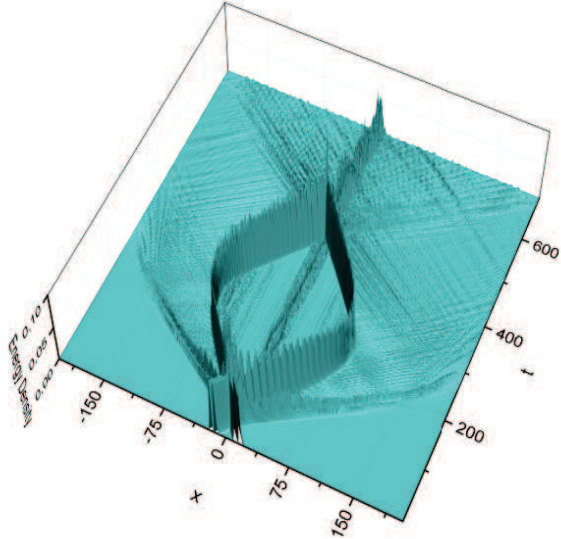}
\end{center}
\vspace{-0.5cm}
\caption{Collision of two-kinks defects for $u=0.5$. Notice the formation of an oscillon at the origin and two symmetric oscillons that collide after bouncing. The outcome is an oscillon at the origin.}
\end{figure}

\begin{figure}[htb]
\begin{center}
\includegraphics*[width=11.cm,height=9.cm]{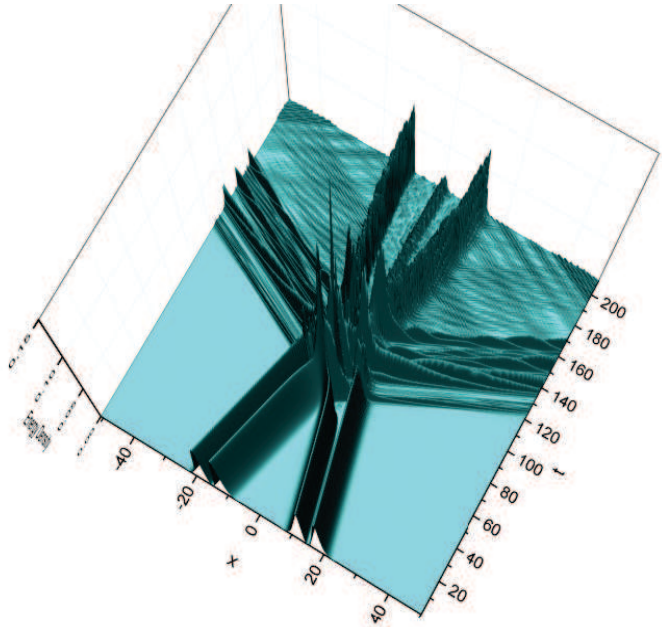}
 \end{center}
\vspace{-0.5cm}
\caption{Illustration of one of the critical configurations obtained for $u=0.1226$. After the collision, two prominent symmetric oscillons remain approximately at rest after receding to a certain distance. We have noticed the formation of an oscillon at the origin. Eventually, both symmetric oscillons coalesce at the origin.}
\end{figure}

The initial impact velocity $u$ is the free parameter that plays a central role in the dynamics of the collision. We have illustrated in Fig. 5 the result of the collision for $u=0.3$ with the three-dimensional plot of the energy density $\rho_\phi$ given by Eq. (\ref{eq5}). Both two-kinks defects collide in a very intricate way resulting in their complete disruption radiating away almost all scalar field. Eventually, a slight amplitude oscillatory bound structure about the origin survives. We have interpreted this bounded state as an oscillon-like structure. We remark that the observed fragmentation of both two-kinks in the first stages of the collision is a general feature regardless the value of the impact velocity and the value of $p \geq 3$. From the computational perspective, the dispersion of a large amount of the scalar field requires collocation points to considerable distances and, for this reason, large map parameter. We remark that this aspect is the leading cause of the relatively high error if compared with the collision in the $\phi^4$ model. In this last model, most of the scalar field is trapped at the origin or escape as traveling waves in localized distributions of energy. 

We have noticed that by setting higher impact velocities a complete fragmentation of both two-kinks. However, the fragmentation gives rise to the appearance of moving symmetric localized distributions of energy, together with a small and oscillating structure about the origin (cf. Fig. 5). We have understood these moving localized distributions energy as oscillon-like structures, from now on we call them oscillons. The formed oscillons recede to a certain distance before returning to collide to form a small amplitude oscillon about the origin. The effect of increasing the impact velocity is to increase the distance both oscillons move apart from each other before bouncing. In particular, these two symmetric prominent oscillons resemble the energy densities of the standard kinks of the $\phi^4$ model. In Fig. 6 we show the collision for $u=0.5$, where the presence of these structures is present. After bouncing the oscillons collide resulting in a small amplitude remnant located at the origin survives. 

We have identified several configurations that appear between the previous outcomes, namely dispersion followed by an oscillon at the origin, and the formation of moving oscillons. These new structures emerge after fine-tuning the impact velocity to some particular values. We have named them \textit{critical configurations} characterized when two symmetric oscillons remain at rest after moving apart each other to some distance. The time the oscillons are in this quasi-stationary phase increases if the impact velocity approaches one of its critical values. We have provided an illustration of one of the critical behaviors in Fig. 7 with $u=0.1226$. It becomes evident the formation of three main oscillons after the collision: one at the origin and two symmetric. These last oscillons move to some distance and stay at rest for some time. Eventually, they coalesce at the origin.

We can further fine-tune the impact velocity inside another interval such that the distance the oscillons recede becomes smaller. As before, the time both oscillons remain frozen depends upon which the fine-tuning to the critical velocity.  We have presented in Fig. 8 two illustrations of the critical configuration with the three-dimensional plots of the scalar field and the energy density. In this case the impact velocity $u=0.39582$. The interaction between both two-kinks occurs at $t \approx 60$, where two oscillons emerge.  It becomes evident that the oscillons remain at a fixed position forming a stationary structure, after moving apart each other. This phase lasts until $t \approx 200$.  
 
\begin{figure*}[htb]
\centering
\includegraphics*[width=8.cm,height=7.cm]{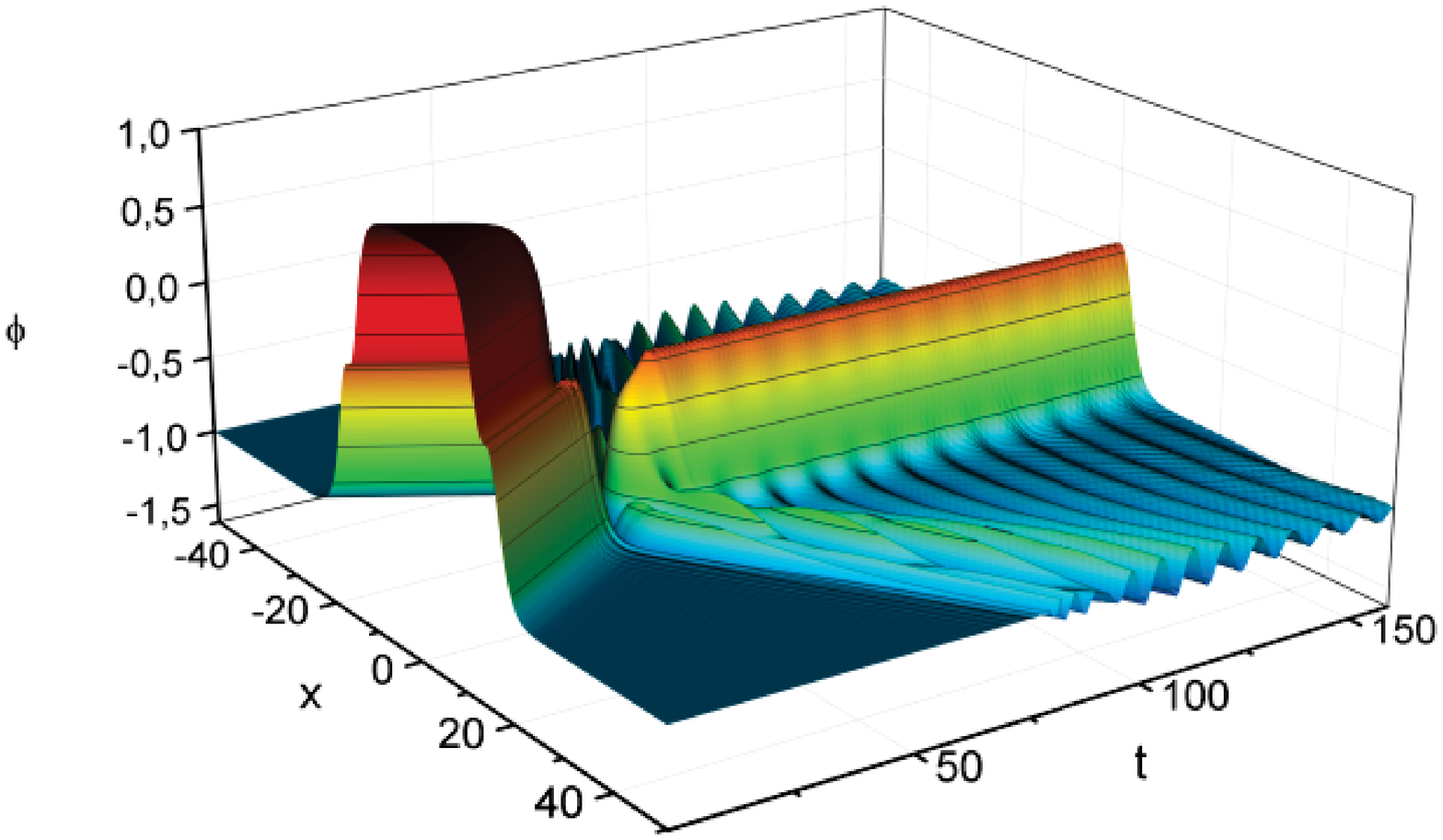}
\hspace{-1.5cm} \includegraphics*[width=8.cm,height=7.cm]{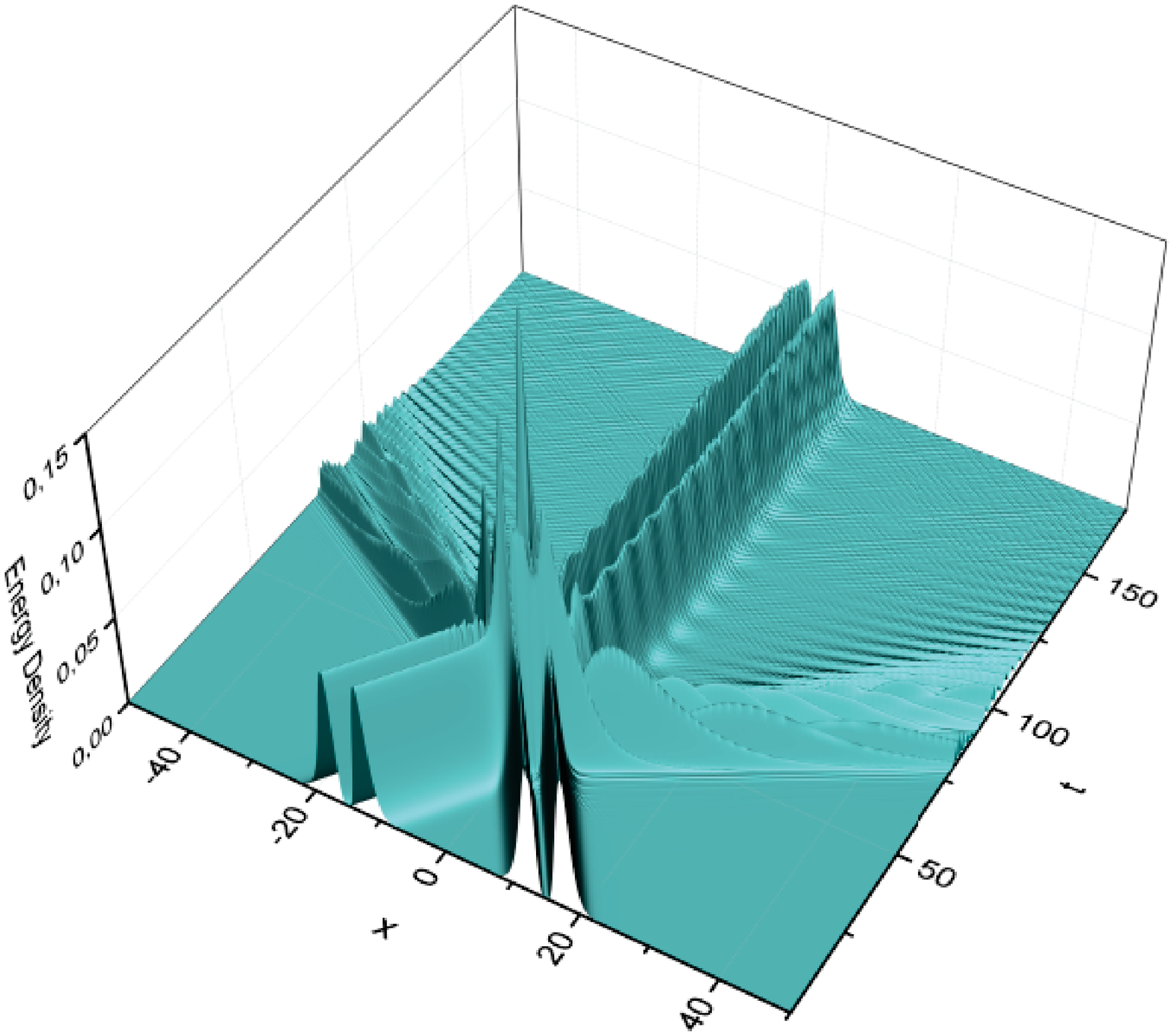}
\includegraphics*[width=5.cm,height=4.cm]{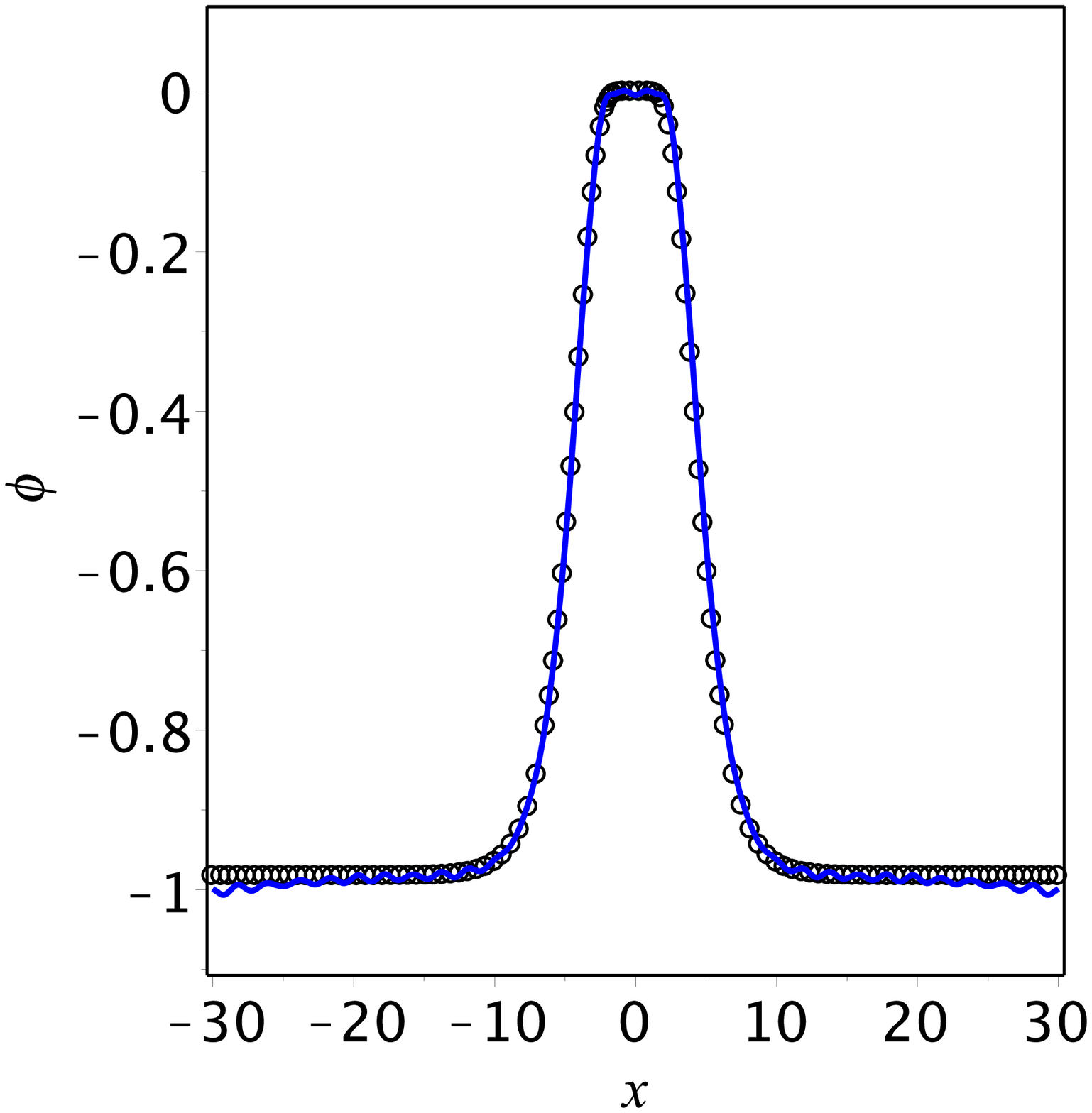}
\hspace{2cm} \includegraphics*[width=5.cm,height=4.cm]{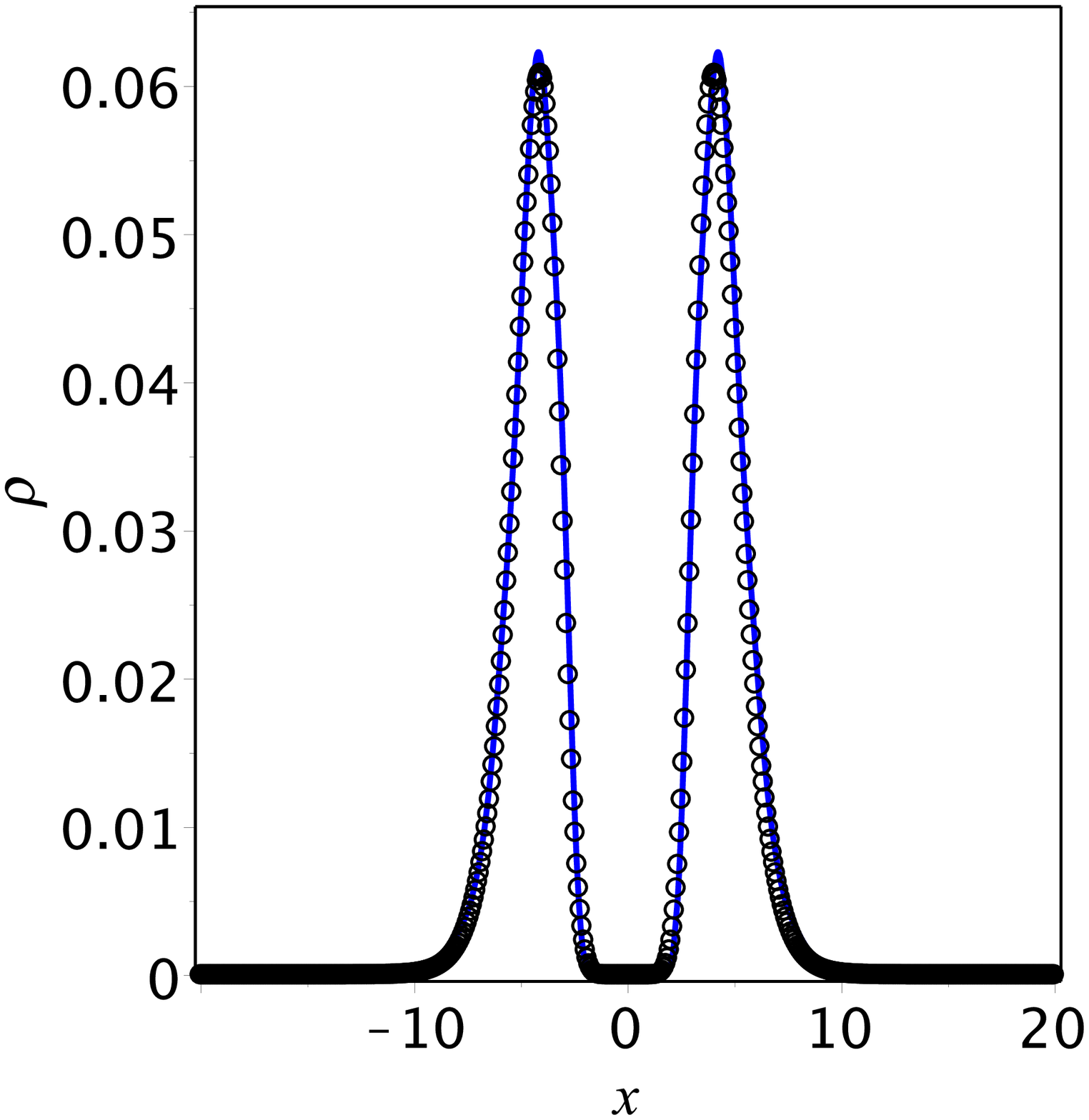}
\caption{Critical configuration formed after the collision of the two-kinks with impact velocity $u=0.39582$. In the upper plots, we have presented the three-dimensional plots of the scalar field and energy density. The stationary configuration is formed at $t \approx 60$ and lasts to $t \approx 200$. In the lower plots, we have depicted the scalar field and energy density profiles evaluated at $t=180$ (blue lines) together with the exact profiles given by Eqs. (\ref{eq20}) and (\ref{eq21}) (black circles). The agreement between the exact and numerical profiles is excellent.}
\end{figure*}

We are compelled to figure out the nature of the stationary configuration shown in Fig. 8. In this direction, we have considered the scalar field profile at $t=180$ during its quasi-stationary phase. The objective is to find an exact solution that reproduces the scalar field and the energy density profiles. We have found that the best candidate that reproduce these numerical patterns is given by,

\begin{equation}
\phi_e(x) = \phi_0 + A_0 \tanh^q\,(b_0 x), \label{eq20}
\end{equation}

\noindent where $\phi_0, A_0$ and $b_0$ are arbitrary constants and $q$ is an even number. The corresponding energy density reads as,

{\small
\begin{eqnarray}
\rho_e(x) = \frac{1}{2} \left(\frac{\partial \delta \phi_e}{\partial x}\right)^2 + \frac{1}{2}b_0^2q^2 \delta \phi_e^2 \left[\left(\frac{\delta \phi_e}{A_0}\right)^{-1/q}-\left(\frac{\delta \phi_e}{A_0}\right)^{1/q}\right]^2, \nonumber \\
\label{eq21}
\end{eqnarray}
}

\noindent where $\delta \phi_e=\phi_e-\phi_0$. The best fit of the numerical scalar field profile at $t=180$ as shown in Fig. 8 is obtained with the following parameters: $\phi_0=0$, $A_0=-0.9833$, $b_0=0.340$ and $q=8$. By inserting these parameters into the energy density (\ref{eq21}) we have also reproduced the numerical profile of the energy density (cf. Fig. 8).
  
\begin{figure*}[htb]
\centering
\includegraphics*[width=5.cm,height=4.cm]{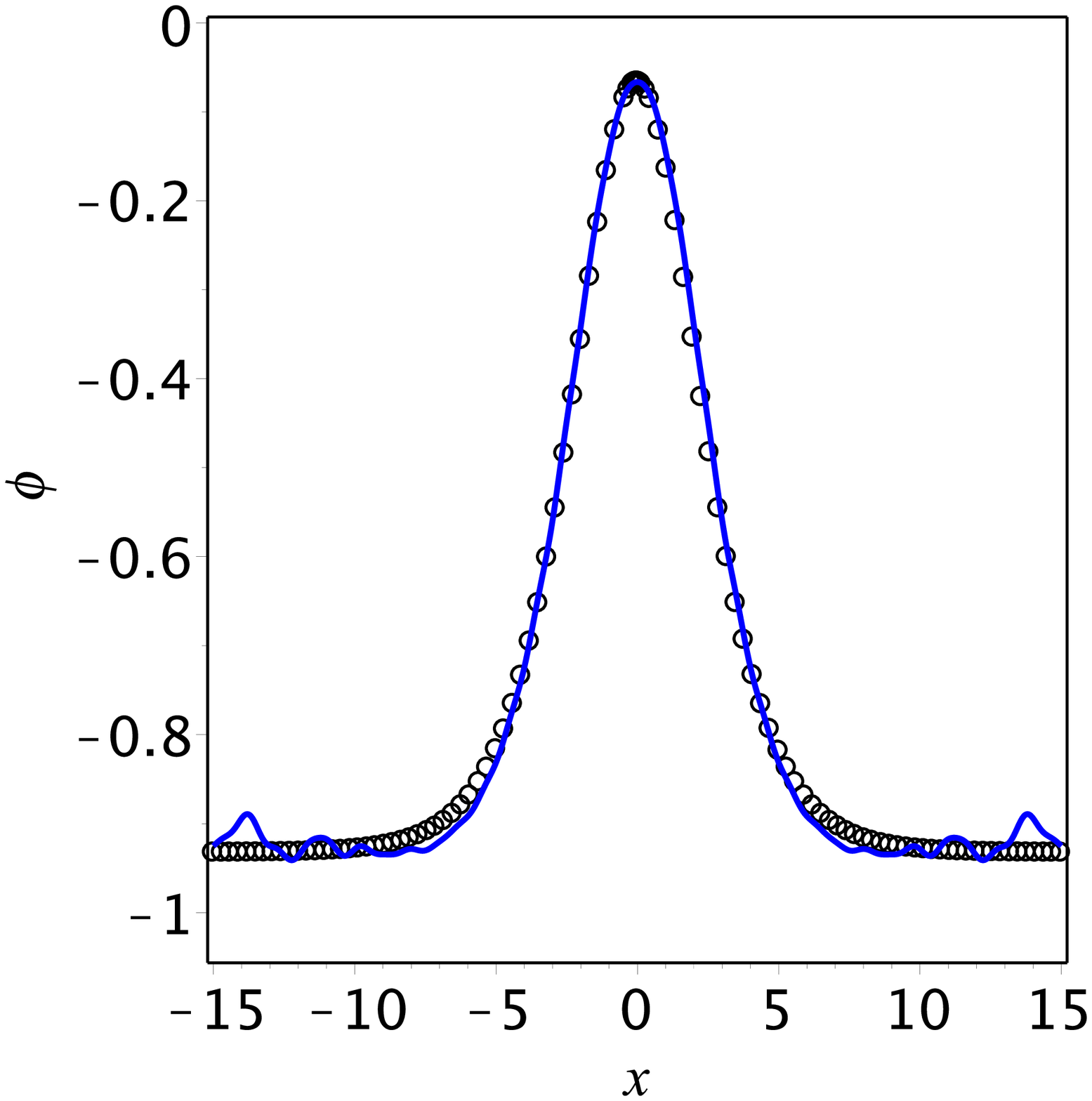} \includegraphics*[width=5.cm,height=4.cm]{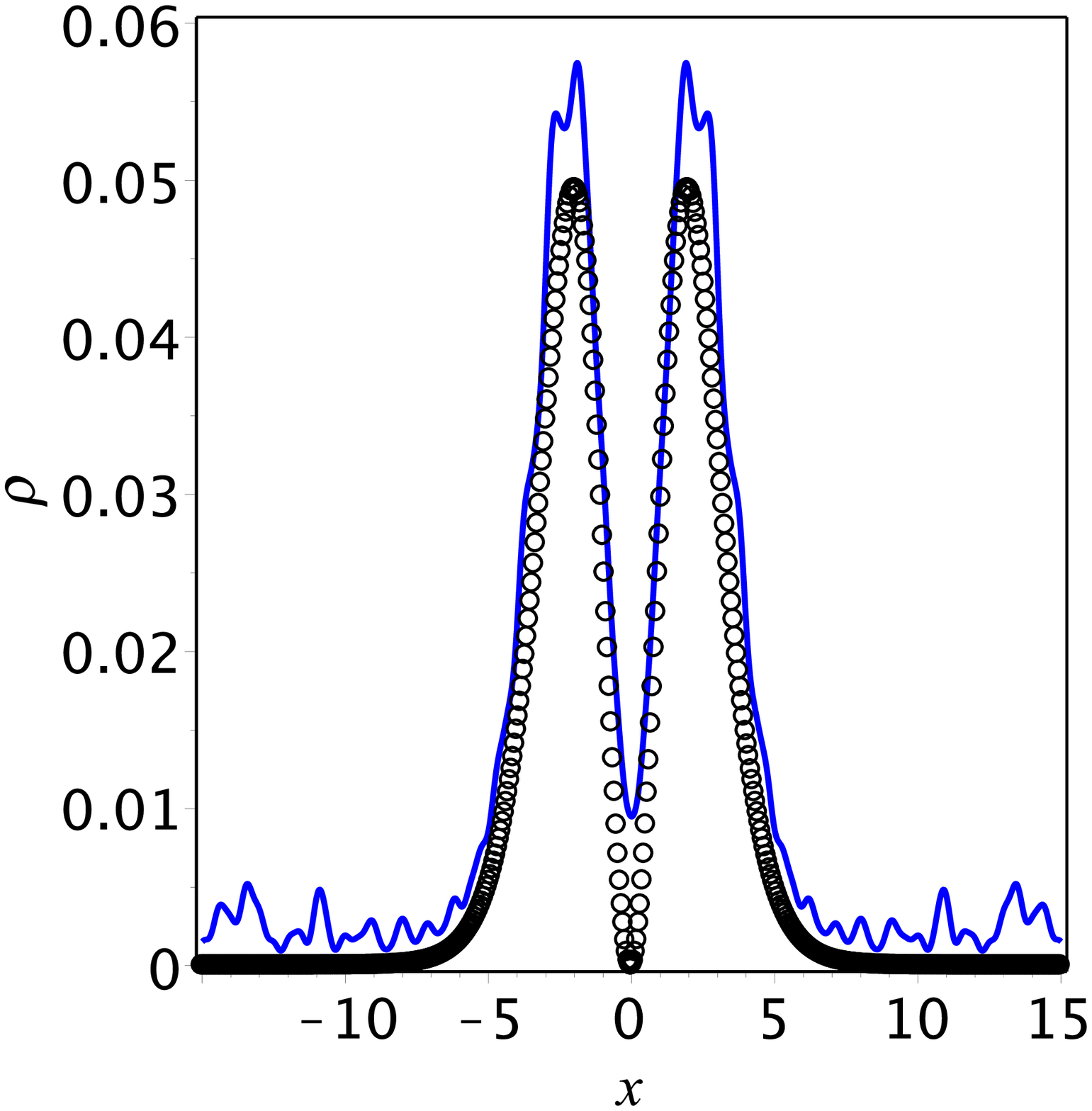}\\
\includegraphics*[width=5.cm,height=4.cm]{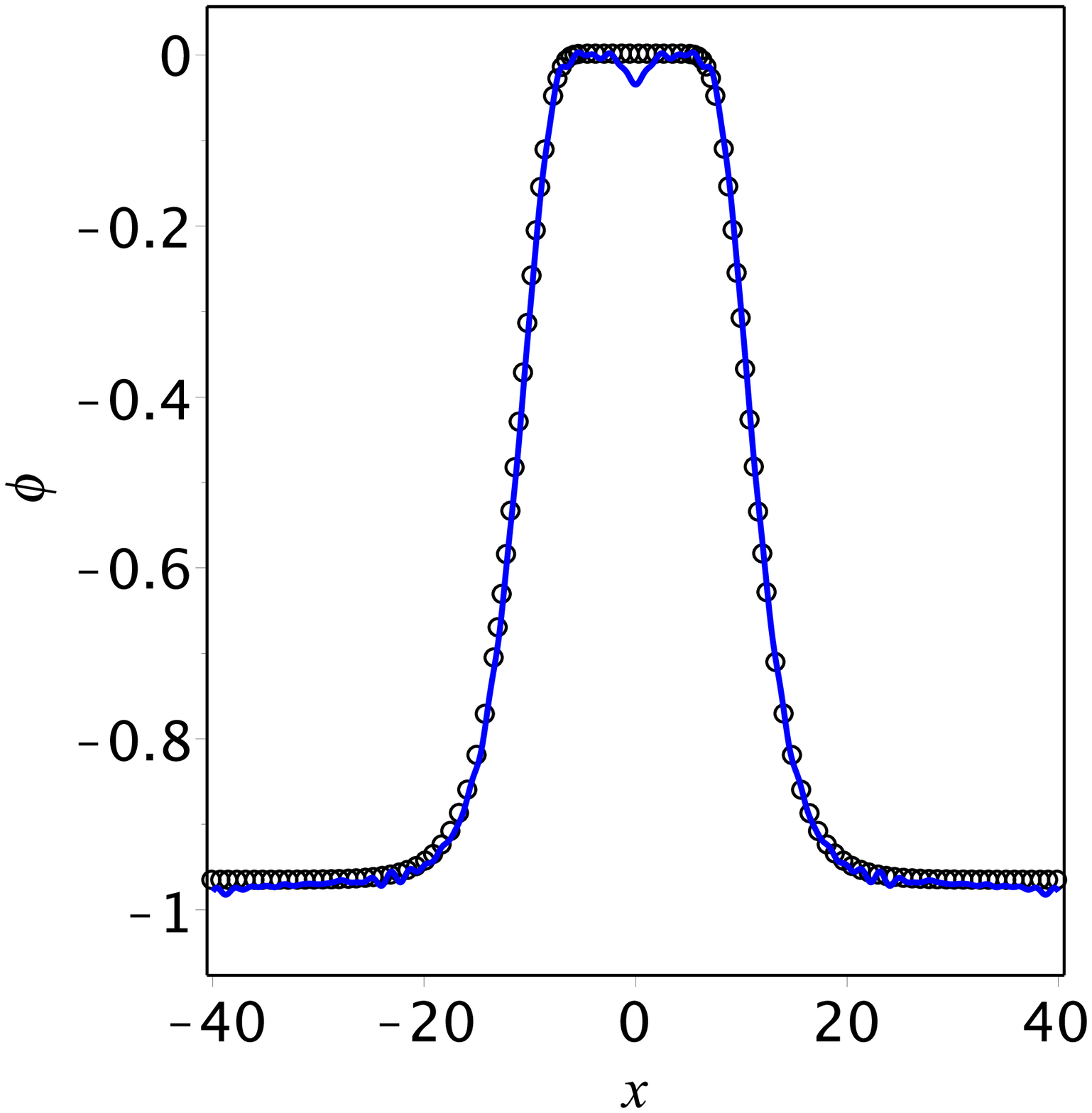} \includegraphics*[width=5.cm,height=4.cm]{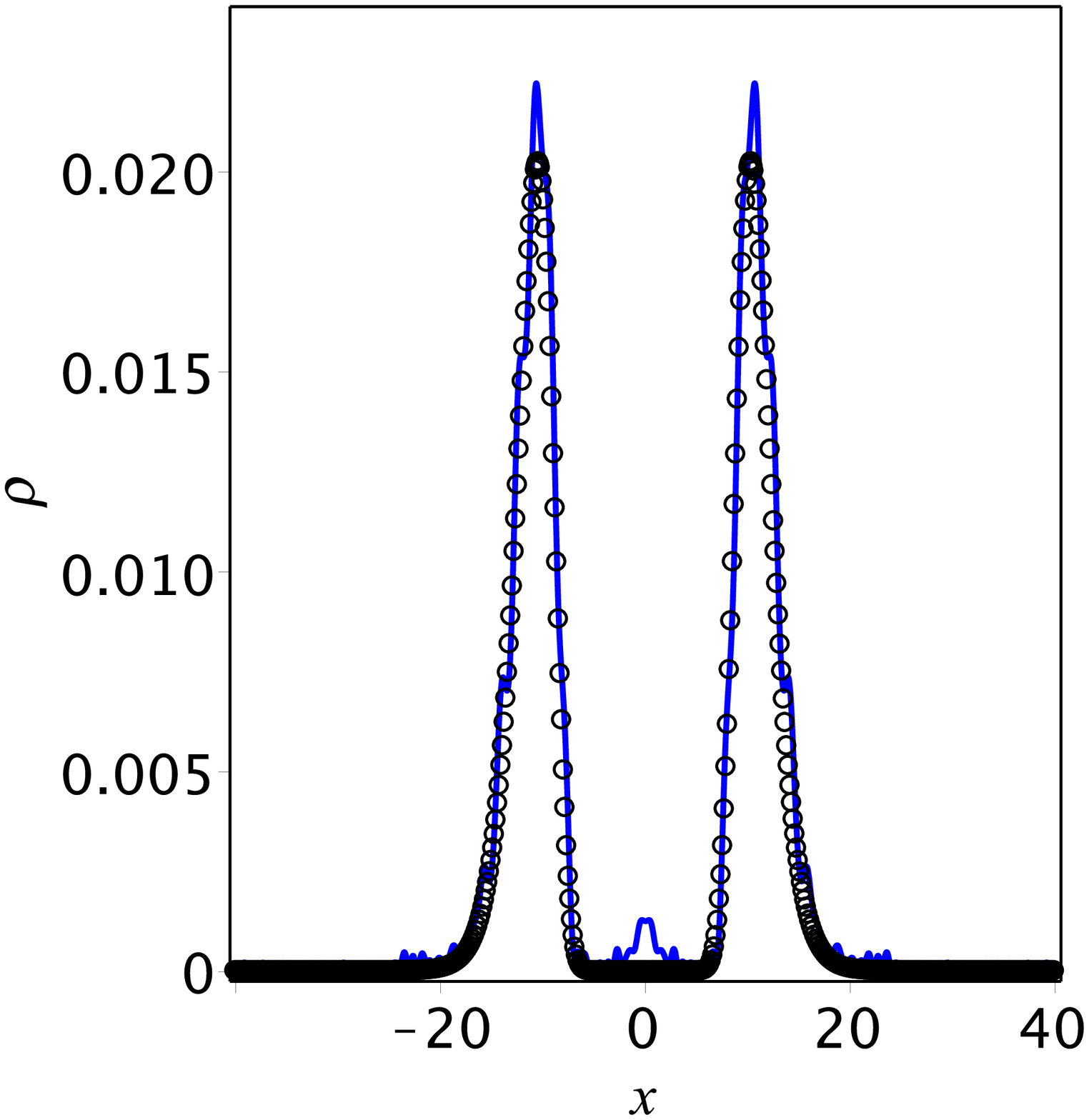}\\
\includegraphics*[width=5.cm,height=4.cm]{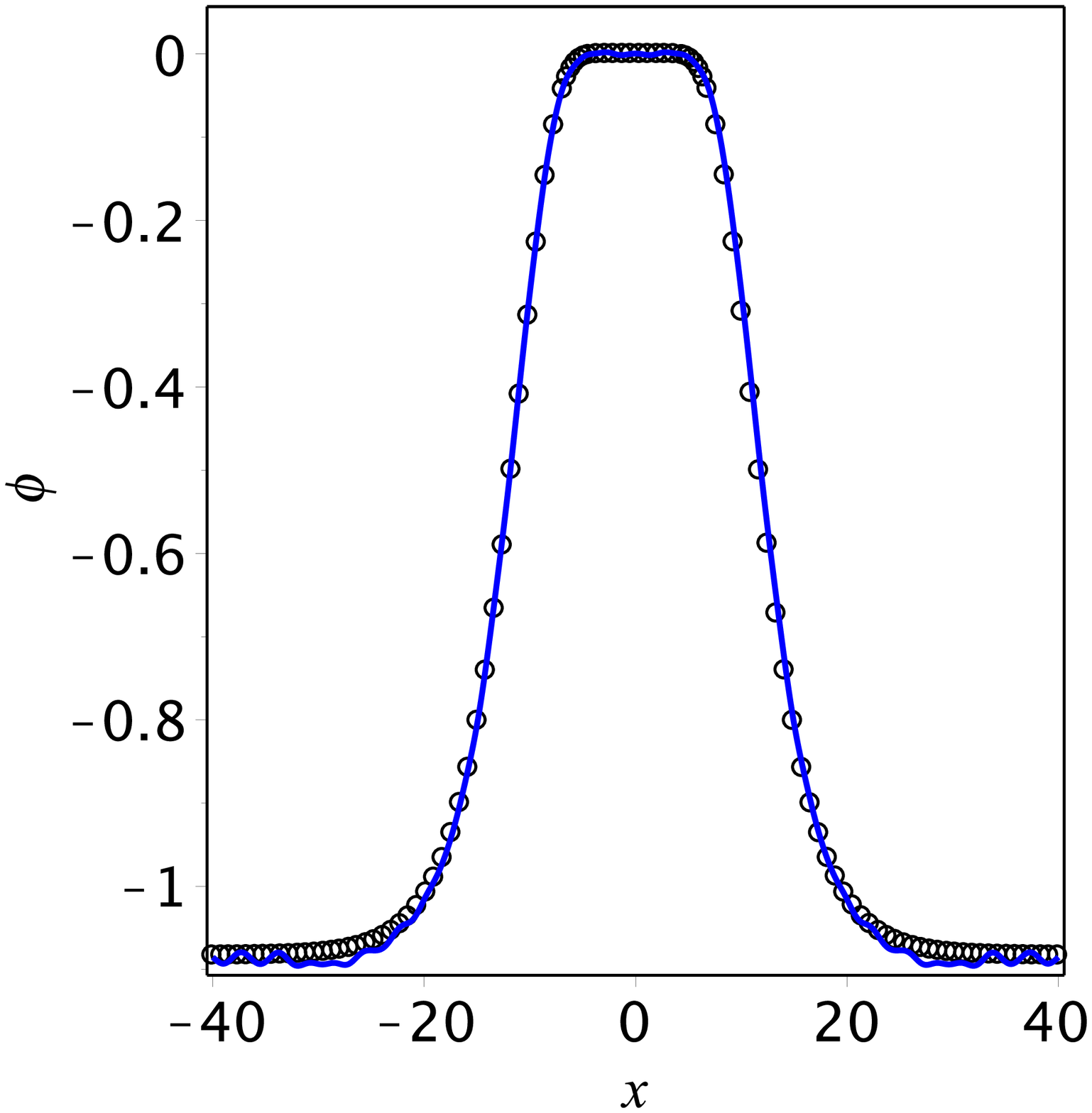} \includegraphics*[width=5.cm,height=4.cm]{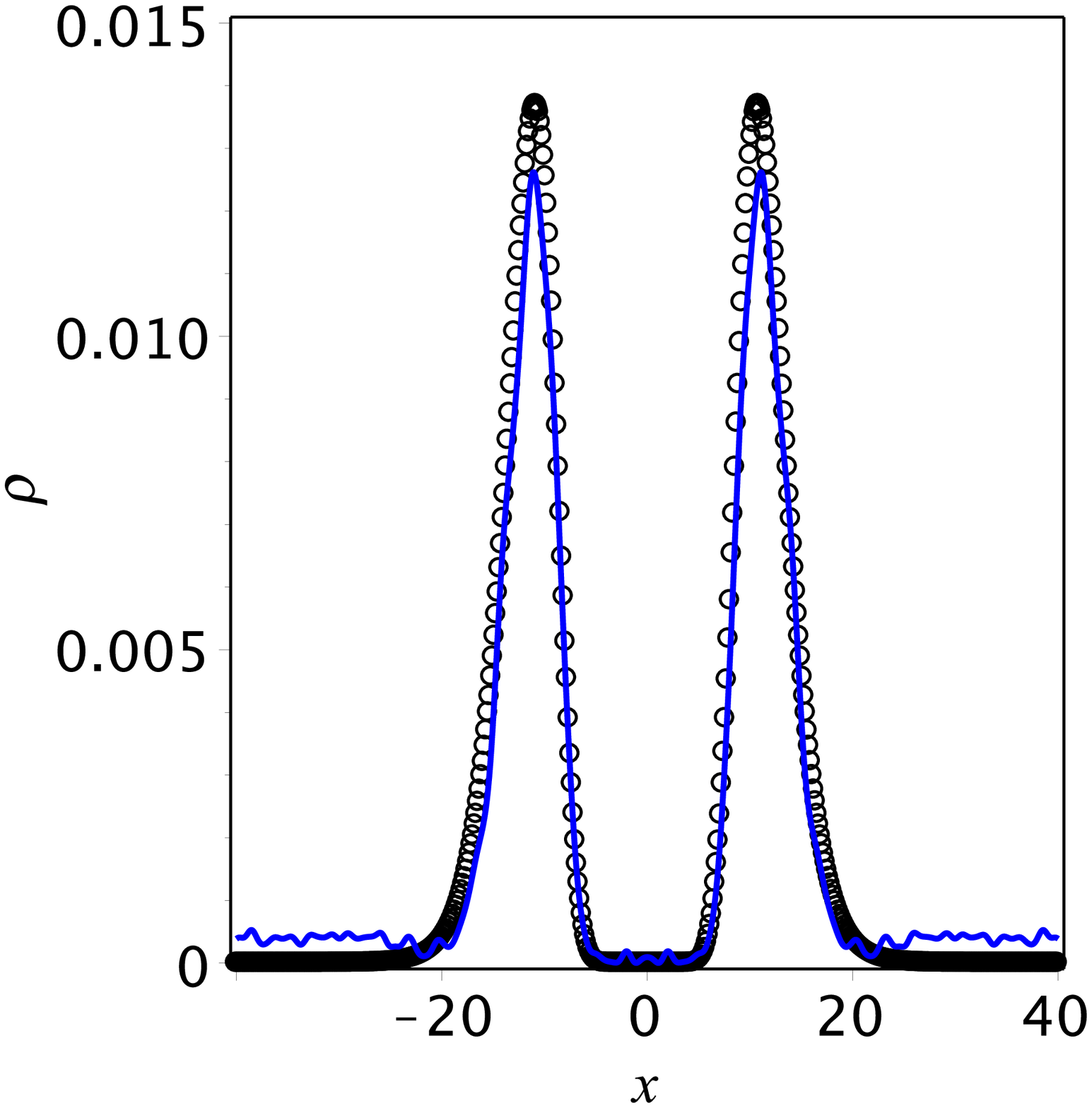}
\caption{Numerical profiles of the scalar field and energy density (blue lines) of the critical configuration together with the corresponding exact profiles given by Eqs. (\ref{eq20}) and (\ref{eq21}) (black circles). From up to down the numerical profiles result after the collision of a two-kink pair in the cases of $p=3$ at $t=135.8$, $p=5$ at $t=350$ and $p=7$ at $t=400$, respectively. The impact velocities are $u=0.1150, 0.11865$ and $u=0.160$. The lumplike critical solutions have the following parameters: (i) $\phi_0 \simeq =-0.06656$, $A_0 \simeq -0.8666$, $b_0 \simeq 0.3333$ and $q=2$; (ii) $\phi_0=0$, $A_0 \simeq -0.9667$, $b_0=0.20$ and $q=32$; and (iii) $\phi_0 \approx 4.8 \times 10^{-4}$, $A_0 \simeq -1.0867$, $b_0 \simeq 0.1466$ and $q=12$.}
\end{figure*}

We remark that the solution (\ref{eq20}) is a slight generalization of the static solution (\ref{eq3}). As we have mentioned the case $q=2$ represents an unstable topological lump-like defect. However, one can verify that the remaining cases, namely $q=4,6,8...$, possess the same properties of a lump-like defect \cite{lump_properties,unusual_lumps}, and consequently we may consider these cases as representing lump-like defects. The numerical experiments indicate that a lump-like defect with $q=8$ emerges as a critical configuration after the collision of two topological two-kinks defects ($p=3$) with impact velocity $u=0.39582$. As far as we are concerned, it is the first time an unstable topological lump-like defect is formed as the result of a collision of a topological defect pair. 

Further numerical experiments have shown the appearance of critical configurations in several intervals of the impact velocity. As a consequence, we conjecture that they are indeed lump-like solutions given by Eq. (\ref{eq20}) with distinct values of $q$. For instance, by adjusting the impact velocity to $u=0.1150$, we have obtained that the critical configuration is approximately represented by a lump-like solution with $q=2$ (cf. Fig. 9). We have noticed in this case that the critical solution survives during a small interval (from  $t \approx 125$ to $t \approx 150$). Also, the value of $\phi_0$ is distinct from zero and can vary with time. It means that the critical configuration can oscillate about the exact static solution (\ref{eq20}). 

We have found critical configurations when a two-kinks pair collide in the case $p >3$. After setting $p=5$ (in this case $x_0=20$) and impact velocity $u=0.11865$, we the lump-like solution has $q=32$ and reproduces the critical configuration. Moreover, if $p=7$ ($x_0=30$) we noticed that a critical configuration emerges when $u=0.160$. We have presented  the corresponding numerical and exact profiles of the scalar field and the energy density for these cases in Fig. 9.

The appearance of the critical configurations can be the result of a nonlinear balance of energy transfer between the vibrational and translational modes of the two-kinks. The nonlinear equilibrium is achieved after an appropriate fine-tuning the impact velocity. In addition, the approach to the unstable lump-like solutions (\ref{eq20}) indicates the existence of channels os attraction that is a new feature of these unstable solutions. Thus, these solutions might be saddle critical points in the abstract space of all possible solutions of the Klein-Gordon equation (\ref{eq1}) with the potential given by Eq. (\ref{eq2}). 


\section{Discussion}

In this paper, we have studied the head-on collision of a new class of kinks known as two-kinks defects introduced by Bazeia et al \cite{bazeia}. We can understand the two-kinks  as composed of two standard kinks of the $\phi^4$ model. For this reason, the collision of a pair of two-kinks can be viewed as a collision of four standard kinks. The Klein-Gordon equation was integrated numerically with a simple and efficient algorithm based on the Collocation method. We have presented numerical tests that confirm the accuracy and the rapid convergence of the algorithm.

We have considered the collision of two-kinks defects for odd $p > 1$. In general, the outcome is the disruption of the two-kinks regardless the impact velocity. The scalar field is almost radiated away leaving behind a small amplitude oscillon at the origin. However, depending on the impact velocity there are two possibilities. The first is a direct formation of a small oscillon at the origin with the emission of almost all scalar field. The second is the formation of outward moving oscillons that eventually bounce and collide to form the final oscillon at the origin. 

The new feature exhibited by the numerical experiments is the presence of several structures that interpolate the above intermediate outcomes. We have called these structures as critical configurations. In these configurations, two symmetric oscillons remain at rest for some time after moving apart to each other. The time the oscillons evolve in this quasi-stationary phase depends on the fine-tuning of the critical velocity to some values. Most importantly, we have identified all critical configurations as belonging to a family of lump-like defects of topological nature and described by Eq. (\ref{eq20}). It constitutes a new feature arising from the collision of a pair of topological defects, in this case, two-kinks defects with $p \geq 3$. 

We remark that a more thoroughly detailed numerical and analytical analysis are necessary for a complete understanding of the appearance of unstable lump-like defects. We have suggested that such a feature might be a consequence of a nonlinear equilibrium of the energy transfer between the vibrational and translational modes of the two-kinks. In this instance, the determination of the internal modes of two-kinks turns to be necessary. In Ref. \cite{bazeia} we found the zeroth mode, and the remaining are obtained after solving the following Schrodinger-like equation for the eigenmodes $\chi_n(x)$, 

\begin{equation}
\left[-\frac{d^2}{dx^2} + \left(\frac{d^2 V}{d \phi^2}\right)_K\right] \chi_n=\omega_n^2\chi_n,
\end{equation}

\noindent with an unusual property of the effective potential diverges at the origin. Another interesting consequence is the characterization of the family of solutions (\ref{eq20}) as saddle-points in the abstract space of solutions of the Klein-Gordon equation (\ref{eq1}). 

\acknowledgments
The authors acknowledge the Brazilian agencies CNPq and CAPES for financial support. We are also grateful to Prof. Bazeia for useful discussions about the two-kinks defects.


\begin{thebibliography}{99}
\bibitem{lump_properties} These properties are: (i) $\lim_{x \rightarrow \pm \infty} \phi = c$, $c$ is a constant, (ii) $\lim_{x \rightarrow \pm \infty} \frac{d\phi}{dx} = 0$, (iii) $\frac{d\phi}{dx} = \pm \sqrt{2 V}, x > 0$ and $\frac{d\phi}{dx} = \mp \sqrt{2 V}, x < 0$. See also Ref. \cite{new_lumplike}.
\end{thebibliography}
\end{document}